# APPLICATION OF GENERALIZED METHOD OF EIGENOSCILLATIONS TO PROBLEMS OF NANOPLASMONICS

M. I. Andriychuk[1], B. Z. Katsenelenbaum[2], V. V. Klimov[3], N. N. Voitovich[1]

[1] Pidstryhach Institute for Applied Problems in Mechanics and Mathematics
NASU, Lviv, Ukraine
e-mail: voi@iapmm.lviv.ua
[2] Nahariya, Israel
[3] P. N. Lebedev Physical Institute, RAS, Moscow, Russia

**Abstract**. A version of generalized eigenoscillation method is applied to the problem about resonant effects in metallic nanoparticles. An approach is proposed, that permits to avoid calculating all higher eigenoscillations except the resonant one. An algorithm for determination of the resonant eigenoscillation, based on the Galerkin procedure, is described in details for the case of bodies of revolution. Model numerical results are presented.

## 1. Introduction

Nanoplasmonics is one of the most intensively developing fields of investigations in the modern physics and its applications. This is confirmed by a great number of publications in the related topics (see, i.e., the recent books [1-6]). It deals, in particular, with the devices which are analogous to those used in the high-frequency electrodynamics and radio engineering. The devices of such a type are resonators, waveguides and antennas. Some theoretical and numerical methods developed for investigation of these devices can be also applied to their plasmonic analogs.

In [6] it was proposed to apply one of the versions of the generalized eigenoscillation method (GEOM) [7], the so-called $\varepsilon$-method, to investigation of the resonant effects in nanoparticles and nanostructures. This variant was firstly proposed in [8]. One of the numerical methods which can be applied for solving homogeneous problems of GEOM is the variational technique. In [9] such technique was developed for bodies of revolution. However, it is not suitable enough for application to the $\varepsilon$-method because it uses the derivatives of higher order, what is not comfortable for numerical calculations. Certain difficulties are connected with the fact that the solutions with negative real part of dielectric permittivity are interesting in nanoplasmonics.

In this paper a numerical scheme designed for solving the nonhomogeneous problems of GEOM is briefly described, and a technique based on the Galerkin approach is applied for solving corresponding homogeneous problems in the case of bodies of revolution. Numerical results for a model problem are presented.

## 2. General scheme of algorithm

In general case, GEOM represents the solution of nonhomogeneous (diffraction) problem near the resonance, as follows

$$\vec{E} = \vec{E}_0 + A_1 \vec{E}_1 + \vec{\Sigma}, \qquad (1)$$

where $\vec{E}$ is the sought electric field, $\vec{E}_0$ is the incident field, $\vec{E}_1$ and $A_1$ are the electric field of the resonant eigenoscillation and its amplitude, respectively, $\vec{\Sigma}$ is the summarized contribution of all other oscillations into the diffracted field. Due to orthogonality of the generalized eigenoscillations (see [7], (1.187)),

$$\int_{V^+} \vec{\Sigma} \cdot \vec{E}_1 dV = 0. \qquad (2)$$

In contrast to [7], where the eigenvalues and fields of nonresonant eigenoscillations are supposed to be known (or if not, their contributions are neglected), we propose to find $\vec{\Sigma}$ directly, without using this information. It can be determined by modification of the linear operator of initial problem, removing from this operator a term generated by the eigenoscillation $\vec{E}_1$. This way provides fulfilling condition (2).

### 3. Homogeneous problem for body of revolution

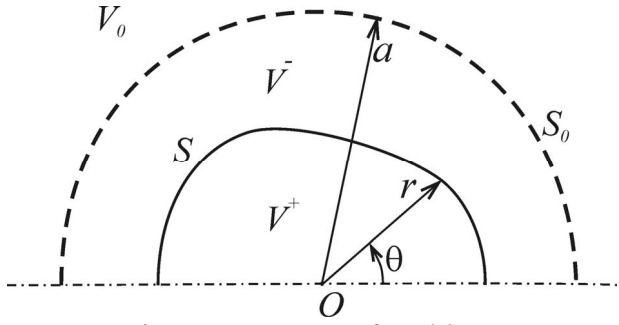

Fig. 1. Geometry of problem.

Here we confine ourselves by the case of particle as a body of revolution with the generatrix $S$. The field components depend on the axial coordinate $\varphi$ of the spherical system $(r,\vartheta,\varphi)$ as $\exp(im\varphi)$ (the time dependence in taken as $\exp(i\omega t)$. The homogeneous problem can be formulated for each value of $m$ separately.

We use the Debye potentials $u(r,\vartheta,\varphi)$, $v(r,\vartheta,\varphi)$ [10] for representation of the eigenoscillation fields in different areas. In order to avoid the collisions connected with the Rayleigh hypothesis, we include into consideration an auxiliary sphere $S_0$ of radius $a$ centered in the coordinate origin and enclosing the boundary $S$ (Fig.1). The coordinate representations of the potentials in different areas are

$$u^0 = \sum A_n^0 P_n^m(\theta) h_n^{(2)}(kr) \exp(im\varphi), \qquad (3a)$$

$$v^0 = \sum C_n^0 P_n^m(\theta) h_n^{(2)}(kr) \exp(im\varphi), \qquad (3b)$$

$$u^- = \sum \left[ A_n^- j_n(kr) + B_n^- y_n(kr) \right] P_n^m(\theta) \exp(im\varphi), \qquad (3c)$$

$$v^- = \sum C_n^- P_n^m(\theta) j_n(kr) \exp(im\varphi) + D_n^- P_n^m(\theta) y_n(kr) \exp(im\varphi), \qquad (3d)$$

$$u^+ = \sum A_n^+ P_n^m(\theta) j_n(kr\sqrt{\varepsilon_1}) \exp(im\varphi), \qquad (3e)$$

$$v^+ = \sum C_n^+ P_n^m(\theta) j_n(kr\sqrt{\varepsilon_1}) \exp(im\varphi). \qquad (3f)$$

Here upper indices 0, +, - coincide with those in notations of the respective areas $V^0, V^-, V^+$; $j_n$, $y_n$, $h_n^{(2)}$ are the Bessel, Neumann and second kind Hankel spherical functions, respectively, $k = 2\pi\omega$ is he wavenumber, $P_n^m(\theta)$ are the associated Legendre functions, $\varepsilon_1$ is the generalized eigenvalue of the problem; the summation is made from $m$ to $\infty$.

The above potentials describe the fields satisfying the homogeneous Maxwell equations in $V^0, V^-$ with $\varepsilon = 1$ and in $V^+$ with $\varepsilon = \varepsilon_1$, as well as the radiation conditions at $r \to \infty$. In the whole space $\mu \equiv 1$ is assumed. The eigenvalue $\varepsilon_1$ and coefficients $A_n^0$, $A_n^-$, $B_n^-$, $A_n^+$, $C_n^0$, $C_n^-$, $D_n^-$, $C_n^+$, corresponding to it, should be determined from the conditions of continuity on the field components $E_\varphi$, $H_\varphi$, $E_s$, $H_s$ on the boundary $S$ and auxiliary sphere $S_0$. Using expressions (7.53) from [10], we write these components for the boundary $S$ described by the equation $r = \rho(\theta)$ (for the sphere $S_0$ one should substitute $\rho = a$, $\rho' = 0$), as

$$E_\varphi = \frac{im}{\rho \sin\theta} \frac{\partial(ru)}{\partial r} + ik\rho \frac{\partial v}{\partial \theta}, \tag{4a}$$

$$H_\varphi = \frac{im}{\rho \sin\theta} \frac{\partial(rv)}{\partial r} - ik\rho\varepsilon^\pm \frac{\partial u}{\partial \theta}, \tag{4b}$$

$$E_S = \frac{1}{\sqrt{\rho^2 + \rho'^2}} \left\{ \rho'\{\frac{\partial^2(ru)}{\partial r^2} + k\rho^2\varepsilon^\pm u\} + \frac{\partial^2(ru)}{\partial r \partial\theta} + \frac{kmr}{\sin\theta} v \right\}, \tag{4c}$$

$$H_S = \frac{1}{\sqrt{\rho^2 + \rho'^2}} \left\{ \rho'\left[\frac{\partial^2(rv)}{\partial r^2} + k^2\rho\varepsilon^\pm v\right] + \frac{\partial^2(rv)}{\partial r \partial\theta} + \frac{km\rho}{\sin\theta}\varepsilon^\pm u \right\}, \tag{4d}$$

where $\varepsilon_1^\pm$ is the dielectric permittivity of the area in which the components are calculated: $\varepsilon^\pm = \varepsilon_1$ for $V^+$, $\varepsilon^\pm = 1$ for $V^-, V^0$. After substituting expressions (3) into (4c), (4d) for $S$, the second derivatives $\partial^2 / \partial r^2$ are reduced in the order by means of the Bessel equation for the spherical functions.

The continuity conditions for these components lead to eight linear functional equations with respect to unknown coefficients. Following to the Galerkin method, we discretize these equations, multiplying each of them by the functions $\psi_q(\theta) = \sin^2\theta \cdot P_q^m(\theta)$, $q = m, m+1, ....$, and integrate over $S$ and $S_0$. Due to the orthogonality of the functions $P_n^m(\theta)$, a part of the terms, corresponding to the conditions on $S_0$, is vanished.

Retaining a finite number $N$ of terms in the sums of (3) and the same number of functions $\psi_q$, we obtain the homogeneous system of $8N$ linear equations with respect to unknown coefficients. After dividing these coefficients into 2 groups: $X_1 = \left[\{A_n^0\}, \{C_n^0\}, \{A_n^-\}, \{B_n^-\}, \{C_n^-\}, \{D_n^-\}\right]^T$, $X_2 = \left[\{A_n^+\}, \{C_n^+\}\right]^T$, ($T$ is the symbol of transposition), this system can be written in the block-matrix form

$$\left[ \begin{vmatrix} A_{11} & A_{12} \\ A_{21} & A_{22} \end{vmatrix} - \varepsilon_1 \begin{vmatrix} 0 & 0 \\ 0 & C \end{vmatrix} \right] \cdot \begin{vmatrix} X_1 \\ X_2 \end{vmatrix} = 0. \qquad (5)$$

The matrices $A_{11}, A_{12}$ are constant, whereas $A_{21}, A_{22}, C$ depend on the sought eigenvalue $\varepsilon_1$ nonlinearly through the arguments of spherical Bessel functions. Besides, the linear factor $\varepsilon_1$ appears in (5), because it is explicitly contained in (4d).

After simple derivations, (5) becomes the standard form

$$[F(\lambda) - \lambda I] \cdot X_2 = 0, \qquad (6)$$

where $F(\lambda) = \{A_{22} - A_{21} \cdot (A_{11}^{-1} \cdot A_{12})\}^{-1} \cdot C$, $\lambda = 1/\varepsilon_1$, $I$ is the $N \times N$ unit matrix. In this way the matrix dimension is reduced in four times.

In (6) we underline the dependence of the matrix $F$ on $\lambda$ in order to apply the iterative procedure in each step of which the linear eigenvalue problem

$$[F(\lambda^{(p-1)}) - \lambda^{(p)} I] \cdot X_2 = 0 \qquad (7)$$

is solved, with the matrix $F$ taken from the previous iteration. The convergence of this procedure is not justified theoretically; however, it (and fast enough) was observed in model numerical experiments, if the initial approximation was properly chosen. Note, that the first step of similar procedure, applied immediately to (6) with $\varepsilon_1^{(0)} = 0$, is equivalent to the quasistatic approximation as described in [6].

The above algorithm was tested on the model problem on the spherical particle of the radius $a$. In Fig.2 the first three eigenvalues are shown versus the dimensionless parameter $ka$. The dashed lines correspond to the Taylor approximations of these values, given in [6], (6.33). The radial distributions of $E_\varphi$-component of corresponding eigenoscillations for spheres with $ka = ka_1 = 0.3$ and $ka = ka_1 = 1$ are presented in Fig.3.

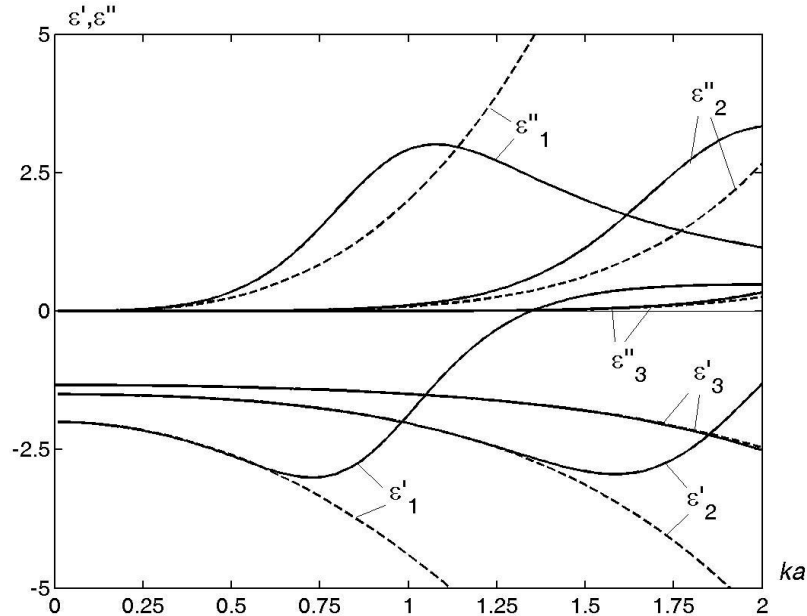

Fig. 2. Eigenvalues of $\varepsilon$-method for spherical particles; $m = 1$.

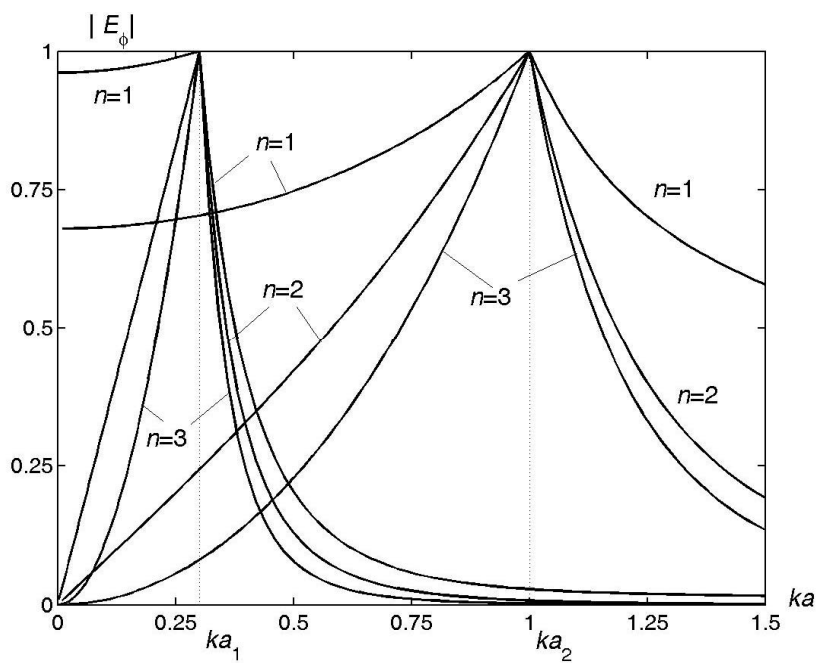

Fig. 3. Field distribution of eigenoscillations for spherical particles; $m = 1$.

V. V. Klimov thanks the Russian Foundation for Basic Research (grants ##11-02-91065, 11-02-92002, 11-02-01272) for partial financial support of this work.